\documentstyle[12pt,a4,psfig]{article}
\begin{document}
\setlength{\oddsidemargin}{-1cm}
\setlength{\evensidemargin}{-1cm}
\newcommand{\nc}{\newcommand}
\nc{\beq}{\begin{equation}}
\nc{\eeq}{\end{equation}}
\nc{\bea}{\begin{eqnarray}}
\nc{\eea}{\end{eqnarray}}
\nc{\ba}{\begin{array}}
\nc{\ea}{\end{array}}
\nc{\nn}{\nonumber}
\nc{\bpi}{\begin{picture}}
\nc{\epi}{\end{picture}}
\nc{\scs}{\scriptstyle}
\nc{\sss}{\scriptscriptstyle}
\nc{\sst}{\scriptstyle}
\nc{\ts}{\textstyle}
\nc{\ds}{\displaystyle}

\nc{\al}{\alpha}
\nc{\be}{\beta}
\nc{\ga}{\gamma}
\nc{\Ga}{\Gamma}
\nc{\de}{\delta}
\nc{\De}{\Delta}
\nc{\ep}{\epsilon}
\nc{\ve}{\varepsilon}
\nc{\eb}{\bar{\eta}}
\nc{\et}{\eta}
\nc{\ka}{\kappa}
\nc{\La}{\Lambda}
\nc{\th}{\theta}
\nc{\Th}{\Theta}
\nc{\ze}{\zeta}
\nc{\p}{\partial}
\nc{\mubar}{\bar{\mu}}

\begin{flushright}
\normalsize
Freiburg-THEP 97/15\\
August 1997\vspace{0.8cm}\\
\end{flushright}

\begin{center}
{\large\bf Perturbative Finite-Temperature Results and
Pad\'{e} Approximants}\vspace{1.2cm}\\
Boris Kastening\vspace{0.4cm}\\
\it Albert-Ludwigs-Universit\"at Freiburg\\
\it Fakult\"at f\"ur Physik\\
\it Hermann-Herder-Stra\ss e 3\\
\it D-79104 Freiburg\\
\it Germany\vspace{1.2cm}\\
\end{center}

\begin{center}
{\bf Abstract}\vspace{0.4cm}\\
\begin{minipage}{16cm}
Pad\'{e} approximants are used to improve the convergence behavior
of perturbative results in massless scalar and gauge field theories
at finite temperature.
\end{minipage}
\vspace{1.2cm}\\
\end{center}

In recent years, computational methods have been developed to analytically
tackle three-loop vacuum diagrams and higher-order contributions of diagrams
with less loops in massless field theories at finite temperature
\cite{PaCo1}-\cite{ZhKa}.
Consequently, the free energy density $F$ at zero chemical potential could
be computed analytically at the $g^5$ level in both massless $g^2\phi^4$
theory \cite{PaSi} (the pressure given there is the negative of the free
energy density) and in massless gauge theories \cite{ZhKa,BrNi}.
In \cite{ZhKa}-\cite{An}, specializations to QED can be found, where the 
result was known before in partially numerical form \cite{PaCo2}.

However, for interesting values of the coupling constant in non-Abelian
gauge theories, the convergence behavior of the perturbative series
is not convincing \cite{ZhKa,BrNi}.
In this brief report, we note that the use of Pad\'{e} approximants 
drastically improves this behavior in both $\phi^4$ and gauge theories.
For the use of Pad\'{e} approximants and other resummation techniques in
other contexts in perturbative field theory and statistical physics,
see, e.g.,\ \cite{SaLi..}, and references therein.

Let us first review those features of the results of \cite{PaSi,ZhKa,BrNi}
which are essential for our analysis here.
The perturbative series for the free energy density in both scalar
and gauge theories has the structure (see the appendix for details)
\beq
\label{fstructure}
F=T^4[c_0+c_2g^2+c_3g^3+(c_{4a}\ln g+c_{4b})g^4+(c_{5a}\ln g+c_{5b})g^5
+O(g^6\ln g)]\,,
\eeq
where $T$ is the temperature and $c_0$, $c_2$, $c_3$, $c_{4a}$, 
$c_{5a}$ are constants, while $c_{4b}$ and $c_{5b}$ have a logarithmic
dependence on $\ln(\mubar/T)$, where $\mubar$ is the renormalization scale
in the modified minimal subtraction scheme ($\overline{\mbox{MS}}$).
In $\phi^4$ theory, $c_{4a}=0$, while in gauge theories $c_{5a}=0$.

As in \cite{ZhKa}, we use the renormalization group to make $g^2$ running,
\beq
\label{runningg}
\frac{1}{g^2(\mubar)}
\approx
\frac{1}{g_T^2}
-\be_0\ln\frac{\mubar}{T}
+\frac{\be_1}{\be_0}\ln\left(1-\be_0g_T^2\ln\frac{\mubar}{T}\right)\,,
\eeq
where $\be_0$ and $\be_1$ are the one- and two-loop coefficients of the
beta function $\be_g$ of $g^2$ (see the appendix for details on $\be_0$
and $\be_1$), and $g_T$ is the coupling constant at temperature $T$.
Then $g$ in (\ref{fstructure}) is replaced by $g(\mubar)$.
In this way we get an idea of the dependence of our result on the
choice of renormalization scale.
We could subsequently expand $F$ in powers of $g_T$ to check that $F$
becomes explicitly independent of $\mubar$ through $g_T^5$.
For this purpose, we would really only need the one-loop coefficient
of $\be_g$.
The reason is that $\be_g$ contains only even powers of $g$,
$\be_g=\be_0g^4+\be_1g^6+\cdots$, since we renormalize as at zero
temperature.
Therefore, from the viewpoint of the renormalization group, the $g^4$
and $g^5$ terms in $F$ are the first corrections to the $g^2$ and $g^3$
terms, respectively.
Numerically, the difference between using $\be_g$ to one or two loops
is insignificant for the examples in non-Abelian gauge theories below,
but keeping the two-loop correction turns out to improve the behavior
of the resummation in $\phi^4$ theory.
The reason why it suffices to use the two-loop beta function in the
examples in this work is that bad behavior of both the perturbative
result and Pade approximants sets in for values of $g^2$ where the
two-loop beta function is still a good approximation.

Now we use Pad\'{e} approximants to reexpress $F$.
For this purpose we pretend that $c_4\equiv c_{4a}\ln g+c_{4b}$ and
$c_5\equiv c_{5a}\ln g+c_{5b}$ are constants in a Taylor series
$F=T^4[c_0+c_2g^2+c_3g^3+c_4g^4+c_5g^5+\cdots]$.
$c_4$ and $c_5$ have different values for each choice of $\mubar$ both
through their direct dependence on $\ln(\mubar/T)$ and through the
running of $g(\mubar)$ in $\ln g$.
Using the approximants [1/2], [2/2] and [2/3] to rewrite $F$ through orders
$g^3$, $g^4$ and $g^5$ (there is no approximant [1/1], since $F$  contains
no term linear in $g$) gives
\bea
F_{[1,2]}
&=&\ts
T^4\frac{c_0^2c_2-c_0^2c_3g}{c_0c_2-c_0c_3 g-c_2^2 g^2}\,,
\\
F_{[2,2]}
&=&\ts
T^4\frac{c_0c_2^2-c_0c_2c_3g+(c_2^3+c_0c_3^2-c_0c_2c_4)g^2}
{c_2^2-c_2c_3g+(c_3^2-c_2c_4)g^2}\,,
\\
\label{f23}
F_{[2,3]}
&=&\ts
T^4\frac{c_0(c_2^3+c_0c_3^2-c_0c_2c_4)
+c_0(-c_2^2c_3-c_0c_3c_4+c_0c_2c_5)g
+(c_2^4+2c_0c_2c_3^2-2c_0c_2^2c_4+c_0^2c_4^2-c_0^2c_3c_5)g^2}
{(c_2^3+c_0c_3^2-c_0c_2c_4)
+(-c_2^2c_3-c_0c_3c_4+c_0c_2c_5)g
+(c_2c_3^2-c_2^2c_4+c_0c_4^2-c_0c_3c_5)g^2
+(-c_3^3+2c_2c_3c_4-c_2^2c_5)g^3}\,.
\eea

Define
\beq
\label{alphat}
\al(T)=g_T^2/(4\pi)
\eeq
and let us look at some specific examples.
Our first case is the small-coupling QCD example from \cite{ZhKa} with
$d_A=8$, $C_A=3$, $n_f=6$, $d_F=18$, $S_F=3$, $S_{2F}=4$, and $\al(T)=0.001$.
As argued in \cite{ZhKa} and as can be seen in Fig.~1a, the perturbative
series for $F$ through $g^5$ is well behaved in this case, with respect
to both convergence for a given renormalization scale $\mubar$ and to the
growing $\mubar$ independence of $F$ towards higher orders.
The Pad\'{e} approximants $F_{[1,2]}$ and $F_{[2,2]}$ are close to the
$g^3$ and $g^4$ results within the expected accuracy (given by the magnitude
of the $g^4$ and $g^5$ corrections, respectively).
However, $F_{[2,3]}$ has a pole, as seen in Fig.~1b.
This pole comes about through a zero of the denominator in 
(\ref{f23}), which, in turn, due to the smallness of $\al(T)$, is
caused by a zero of the first term in the denominator of (\ref{f23}),
$c_2^3+c_0c_3^2-c_0c_2c_4$.
We know that the full result for $F$ is independent of $\mubar$ and 
that consequently this pole is an artifact of the resummation scheme.
We therefore determine its position and residue, explicitly remove 
it and call the resulting function $\bar{F}_{[2,3]}$.
The curve in Fig.~1b for $\bar{F}_{[2,3]}$ is virtually identical to the
$g^5$ result in the perturbative series in Fig.~1a.

Now let us turn to cases where the pure perturbative series needs
improvement.
Fig.~2a represents the perturbative series for the pure SU(2) example
from \cite{ZhKa} with $d_A=3$, $C_A=2$, $n_f=d_F=S_F=S_{2F}=0$, and
$\al(T)=0.03$, while Fig.~2b shows the Pad\'{e} approximants.
Again, we have removed the pole from $F_{[2,3]}$ to define
$\bar{F}_{[2,3]}$ and show both functions.
Clearly, the convergence behavior of the series $F_{[1,2]}$, $F_{[2,2]}$,
$\bar{F}_{[2,3]}$ is drastically improved compared to the purely perturbative
series, particularly around natural choices of $\mubar$, such as $\mubar=T$
or $\mubar=gT$, which up to $g^5$ order are the only mass scales in
finite-temperature non-Abelian gauge theories.
Note also the relative independence from the renormalization scale.

Now turn to the other QCD example in \cite{ZhKa}, namely, $\al(T)=0.1$
with the other parameters being the same as in our first case.
The result is plotted in Fig.~3.
Again, there is much improvement compared to the pure perturbative
series around $\mubar=T\approx gT$, where higher approximants give
subsequently smaller corrections to their predecessors.

As our final example in non-Abelian gauge theories, consider three-flavor
QCD, i.e., $d_A=8$, $C_A=3$, $n_f=3$, $d_F=9$, $S_F=3/2$, $S_{2F}=2$
with $\al(2\pi T)=1/3$ [note that we have to replace $T\rightarrow 2\pi T$
in (\ref{runningg}) and (\ref{alphat}) accordingly].
Up to the fact that we neglect the strange-quark mass and that we have
set all chemical potentials to zero, this is close to the case of the
quark-gluon plasma to be produced at the BNL Relativistic Heavy-Ion
Collider (RHIC).
The result is plotted in Fig.~4.
There seems to be no useful improvement over the perturbative series,
although the range of manifestly bad behavior is shifted towards smaller
values of $\mubar$.

In Fig.~5, we present an example in scalar theory, namely with $\al(T)=0.75$.
Note how, at least for not too big $\mubar$, the Pad\'{e} approximants
fluctuate much less in subsequent orders than their purely perturbative
counterparts.
The fact that we can go to larger couplings in scalar theory than in
non-Abelian gauge theory is easily explained.
For example, for the case of no fermions, the effective expansion parameter
in $F$ is seen to be $C_A\al(T)$.
That is, a larger number of degrees of freedom leads to stronger corrections
to the ideal gas result (unless we try to make fermionic and bosonic
contributions cancel).

Let us make two final remarks.
(i) The use of the approximants [2/1] and [3/2] instead of [1/2] and [2/3]
gives results very similar to those presented here, while approximants
$[m/n]$ with $|m-n|>1$ give typically less improvement.
(ii) Starting at $g^6$ order, another physical scale $g^2T$ enters the
calculation of $F$ in non-Abelian gauge theories \cite{Li..}.
Therefore, it would be interesting to see how inclusion of the $g^6$
term changes our results.
Unfortunately, computation of this term is difficult and requires a
combination of perturbative and nonperturbative techniques \cite{BrNi,Br}.

\subsection*{Acknowledgements}
I am grateful to G.~Jikia for helpful comments and to E.~Braaten for
pointing out an error in the original manuscript.
This work was supported by the Deutsche Forschungsgemeinschaft (DFG).

\appendix

\section*{Appendix}

Here we give the results of \cite{PaSi} and \cite{ZhKa,BrNi}.

With the Euclidean Lagrange density
\beq
{\cal L}=\frac{1}{2}(\p_\mu\phi)^2+\frac{g^2}{4!}\phi^4\,,
\eeq
the free energy density in the $\overline{\mbox{MS}}$ scheme is
\bea
F
&=&
-\frac{\pi^2T^4}{9}
\Bigg\{
\frac{1}{10}-\frac{1}{8}\left(\frac{g}{4\pi}\right)^2
+\frac{1}{\sqrt{6}}\left(\frac{g}{4\pi}\right)^3
\nn\\
&&\hspace{40pt}
-\left(\frac{g}{4\pi}\right)^4
\left[-\frac{3}{8}\ln\frac{\mubar}{4\pi T}
+\frac{1}{4}\frac{\ze'(-3)}{\ze(-3)}
-\frac{1}{2}\frac{\ze'(-1)}{\ze(-1)}
-\frac{1}{8}\ga_E+\frac{59}{120}\right]
\nn\\
&&\hspace{40pt}
+\left(\frac{g}{4\pi}\right)^5
\sqrt{\frac{3}{2}}\left[-\frac{3}{2}\ln\frac{\mubar}{4\pi T}
+\frac{\ze'(-1)}{\ze(-1)}-\frac{1}{2}\ga_E-\frac{5}{4}
+2\ln\left(\frac{g}{4\pi}\sqrt{\frac{2}{3}}\right)\right]
\Bigg\}
+O(g^6\ln g)\,,\hspace{20pt}
\eea
where we have translated the MS result of \cite{PaSi} into 
$\overline{\mbox{MS}}$ using $\mu^2=e^{\ga_E}\mubar^2/(4\pi)$
and where $\zeta$ is Riemann's zeta function and
$\gamma_E$ is the Euler-Mascheroni constant.
The one- and two-loop coefficients in $\be_g$ are
\beq
\be_0=\frac{3}{(4\pi)^2}\,,
\hspace{30pt}
\be_1=-\frac{17}{3(4\pi)^4}\,.
\eeq

In gauge theory with fermions with a single, simple Lie group
consider the Euclidean Lagrange density
\beq
{\cal L} =
\bar{\psi}\gamma_\mu\left(\partial_\mu-igA_\mu^a T^a\right)\psi
+\frac{1}{4}\left(\partial_\mu A_\nu^a -\partial_\nu A_\mu^a
+gf^{abc}A_\mu^b A_\nu^c\right)^2\,,
\eeq
where the $T^a$ are the generators of the group in the fermion
representation.
Let $d_A$ and $C_A$ be the dimension and quadratic
Casimir invariant of the adjoint representation, with
\begin{equation}
\delta^{aa}=d_A\,,
\hspace{20pt}
f^{abc}f^{dbc}=C_A\delta^{ad}\,.
\end{equation}
Let $d_F$ be the dimension of the total fermion representation
({\it e.g.}, 18 for six-flavor QCD), and define $S_F$ and
$S_{2F}$ in terms of the generators $T^a$ for the total fermion
representation as
\beq
S_F =\frac{1}{d_A}{\rm tr}(T^2)\,,
\hspace{20pt}
S_{2F} = \frac{1}{d_A}{\rm tr} [(T^2)^2]\,,
\eeq
where $T^2 = T^a T^a$.
For SU($N$) with $n_F$ fermions in the fundamental representation,
the standard normalization of the coupling gives
\beq
d_A = N^2-1\,,
\hspace{10pt}
C_A=N\,,
\hspace{10pt}
d_F=N n_F\,,
\hspace{10pt}
S_F=\frac{1}{2} n_F\,,
\hspace{10pt}
S_{2F}=\frac{N^2-1}{4N}n_F\,.
\eeq
The the free energy density is given by
\bea
F
&=&
d_AT^4\frac{\pi^2}{9}\Bigg\{
-\frac{1}{5}\left(1+\frac{7d_F}{4d_A}\right)
+\left(\frac{g}{4\pi}\right)^2
\left(C_A+\ts\frac{5}{2}S_F\right)
\nn\\
&&\hspace{50pt}
-48\left(\frac{g}{4 \pi}\right)^3
\left(\frac{C_A+S_F}{3}\right)^{3/2}
-48\left(\frac{g}{4\pi}\right)^4C_A(C_A+S_F)
\ln\left(\frac{g}{2\pi}\sqrt{\frac{C_A+S_F}3}\right)
\nn\\
&&\hspace{50pt}
+\left(\frac{g}{4\pi}\right)^4\Bigg[
C_A^2
\left(\frac{22}{3}\ln\frac{\mubar}{4\pi T}
+\frac{38}{3}\frac{\ze'(-3)}{\ze(-3)}
-\frac{148}{3}\frac{\ze'(-1)}{\ze(-1)}
-4\ga_E+\frac{64}{5}\right)
\nn\\
&&\hspace{100pt}
+C_AS_F
\left(\frac{47}{3} \ln\frac{\mubar}{4\pi T}
+\frac{1}{3} \frac{\ze'(-3)}{\ze(-3)}
-\frac{74}{3} \frac{\ze'(-1)}{\ze(-1)}
-8\ga_E+\frac{1759}{60}+\frac{37}{5}\ln 2\right)
\nn\\
&&\hspace{100pt}
+S_F^2
\left(-\frac{20}{3}\ln\frac{\mubar}{4\pi T}
+\frac{8}{3}\frac{\ze'(-3)}{\ze(-3)}
-\frac{16}{3}\frac{\ze'(-1)}{\ze(-1)}
-4\ga_E-\frac{1}{3}+\frac{88}{5}\ln 2\right)
\nn\\
&&\hspace{100pt}
+S_{2F}\left(-\frac{105}{4}+24\ln 2\right)\Bigg]
\nn\\
&&\hspace{50pt}
-\left(\frac{g}{4\pi}\right)^5
\left(\frac{C_A+S_F}{3}\right)^{1/2}
\Bigg[C_A^2\left(176\ln\frac{\mubar}{4\pi T}
+176 \ga_E-24\pi^2-494+264\ln 2\right)
\nn\\
&&\hspace{175pt}
+ C_A S_F\left (112 \ln\frac{\mubar}{4\pi T}
+112\ga_E+72-128 \ln 2\right)
\nn\\
&&\hspace{175pt}
+ S_F^2\left (-64 \ln\frac{\mubar}{4\pi T}
-64\ga_E+32-128 \ln 2\right)
\nn\\
&&\hspace{175pt}
-144 S_{2F}\Bigg]+O(g^6)\Bigg\}\,.
\eea
The one- and two-loop coefficients in $\be_g$ are
\beq
\be_0=\frac{1}{(4\pi)^2}\left(-\frac{22}{3}C_A+\frac{8}{3}S_F\right)\,,
\hspace{30pt}
\be_1=\frac{1}{(4\pi)^4}\left(
-\frac{68}{3}C_A^2+\frac{40}{3}C_AS_F+8S_{2F}\right)\,.
\eeq

\begin{figure}

\psfig{figure=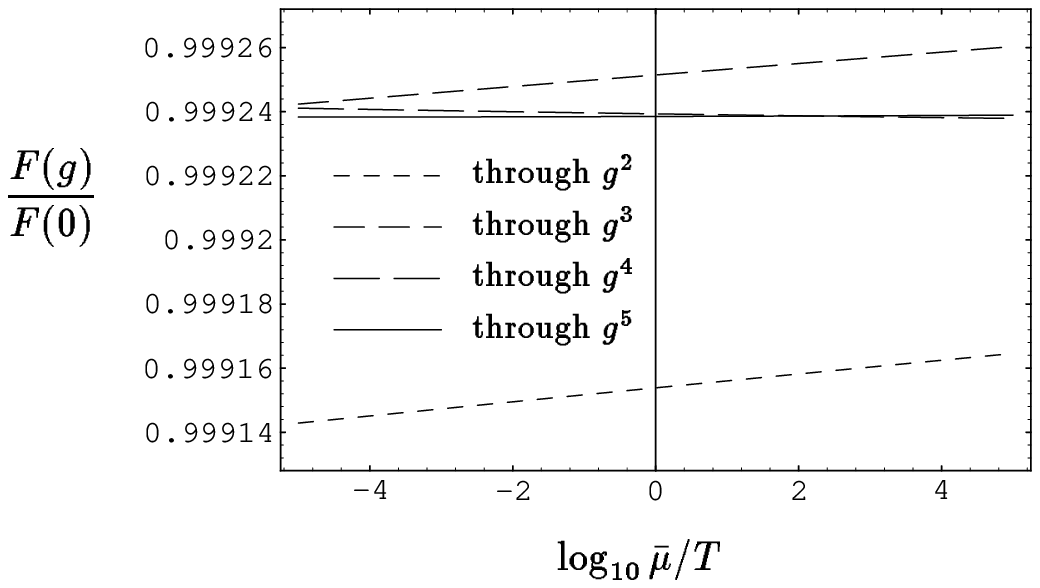,width=5.875in,bbllx=36pt,bblly=482pt,bburx=505pt,bbury=655pt}
\begin{center}
Fig.~1a
\end{center}

\psfig{figure=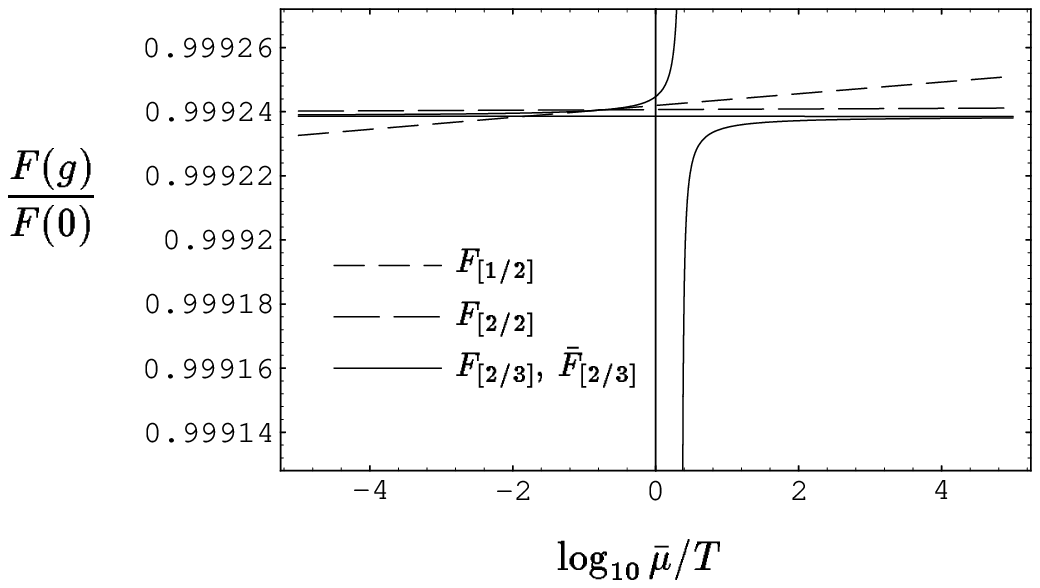,width=5.875in,bbllx=36pt,bblly=482pt,bburx=505pt,bbury=655pt}
\begin{center}
Fig.~1b
\end{center}

\caption{Fig.~1a shows the perturbative series for the free
energy density in units of the ideal gas result $F(g=0)$ for
six-flavor QCD with $\al(T)=0.001$ for a range of choices
of renormalization scale $\mubar$.
The short-dashed, medium-dashed, long-dashed and solid lines are the
results for $F$ including terms through orders $g^2$, $g^3$, $g^4$ and
$g^5$, respectively.
Fig.~1b shows Pad\'{e} approximants instead: The $g^2$ result
has been dropped, while the result through $g^3$ from Fig.~1a has
been replaced by $F_{[1/2]}$, the result through $g^4$ by
$F_{[2/2]}$ and the result through $g^5$ by $F_{[2/3]}$
(solid line with pole) and $\bar{F}_{[2/3]}$ (solid line without pole).}
\end{figure}

\begin{figure}

\psfig{figure=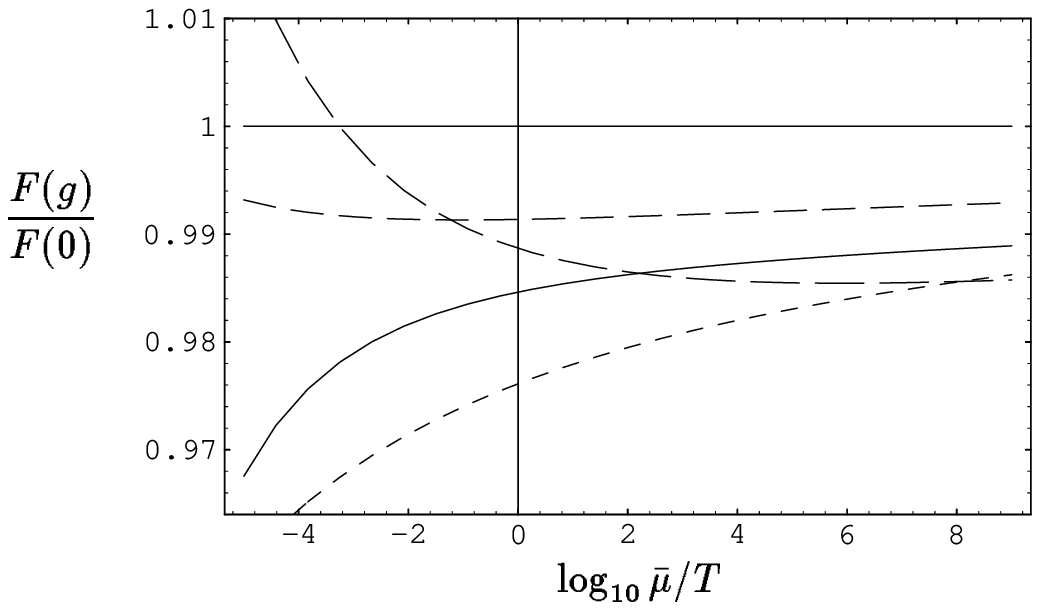,width=5.875in,bbllx=36pt,bblly=482pt,bburx=505pt,bbury=655pt}
\begin{center}
Fig.~2a
\end{center}

\psfig{figure=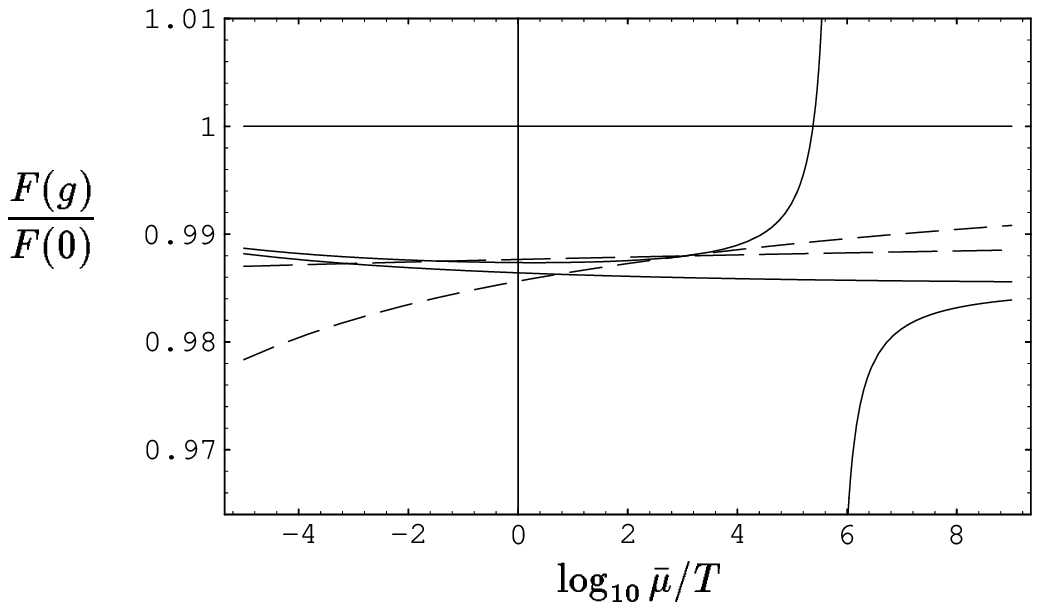,width=5.875in,bbllx=36pt,bblly=482pt,bburx=505pt,bbury=655pt}
\begin{center}
Fig.~2b
\end{center}

\caption{The same as Fig.~1, but for pure SU(2) theory with $\al(T)=0.03$.}
\end{figure}

\begin{figure}

\psfig{figure=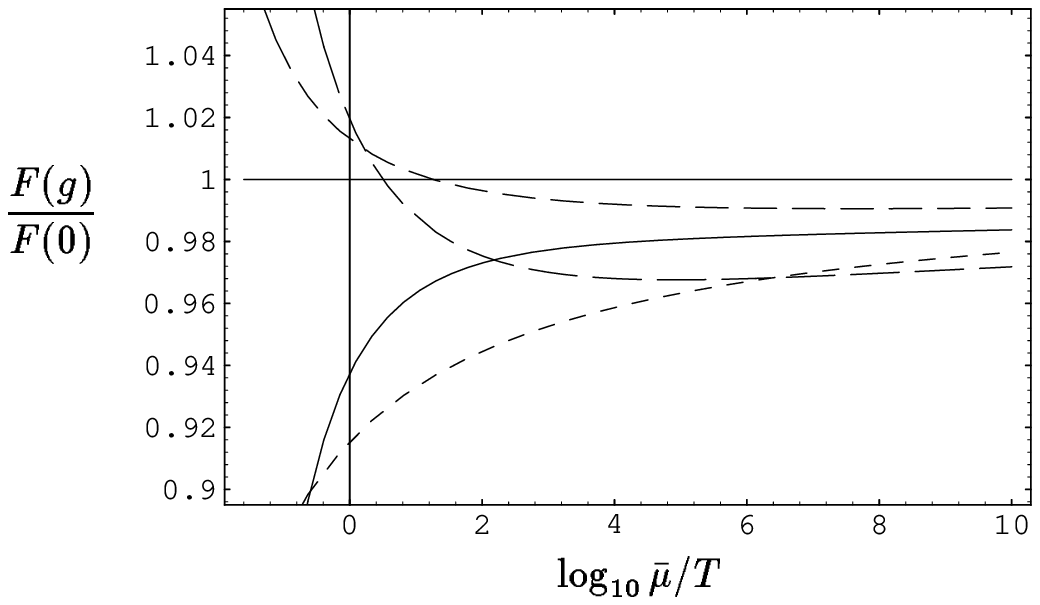,width=5.875in,bbllx=36pt,bblly=482pt,bburx=505pt,bbury=655pt}
\begin{center}
Fig.~3a
\end{center}

\psfig{figure=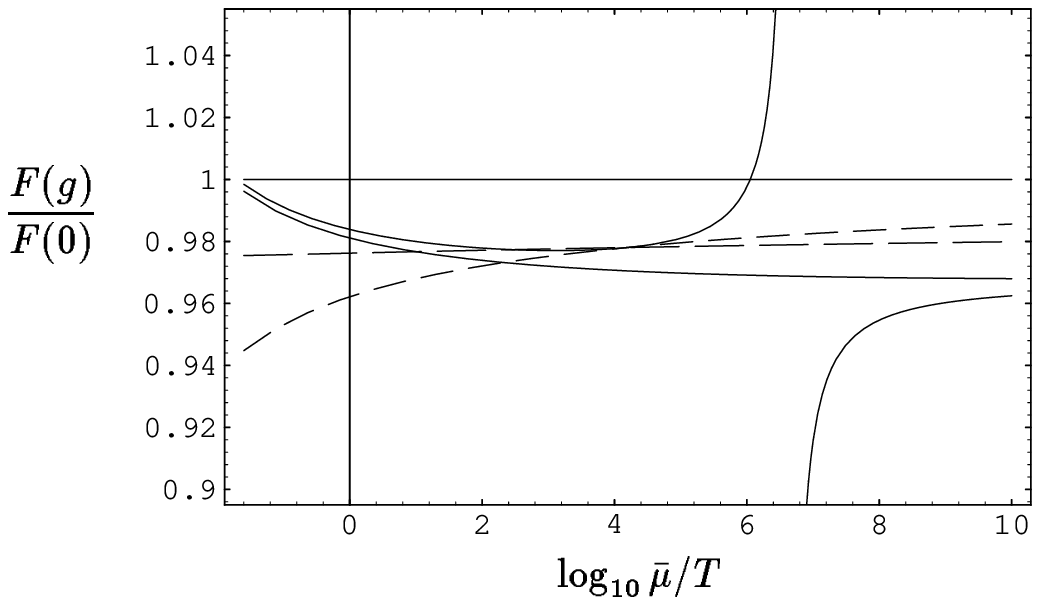,width=5.875in,bbllx=36pt,bblly=482pt,bburx=505pt,bbury=655pt}
\begin{center}
Fig.~3b
\end{center}

\caption{The same as Fig.~1, but with $\al(T)=0.1$.}
\end{figure}

\begin{figure}

\psfig{figure=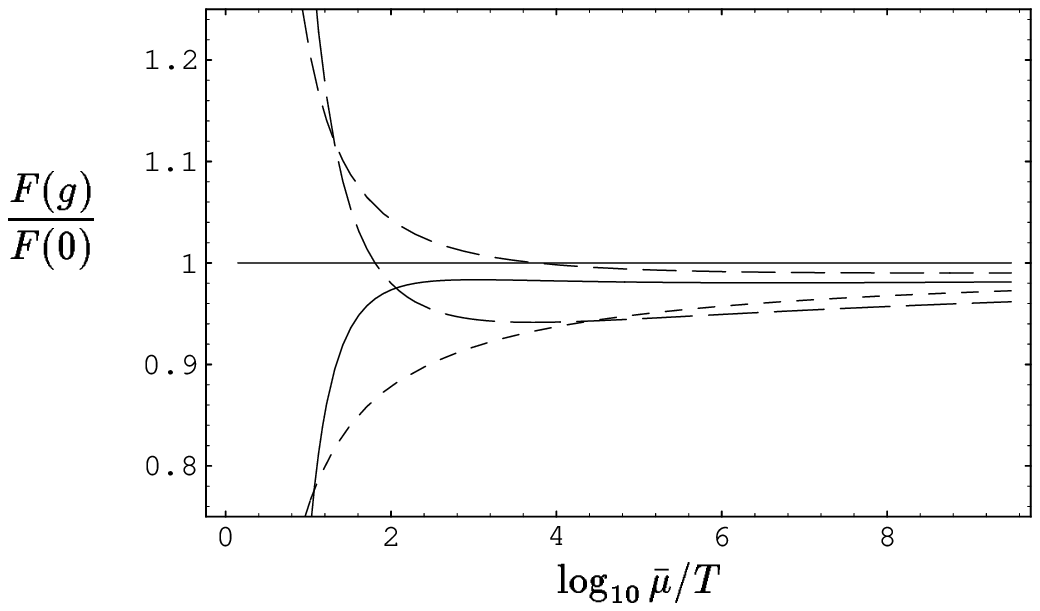,width=5.875in,bbllx=36pt,bblly=482pt,bburx=505pt,bbury=655pt}
\begin{center}
Fig.~4a
\end{center}

\psfig{figure=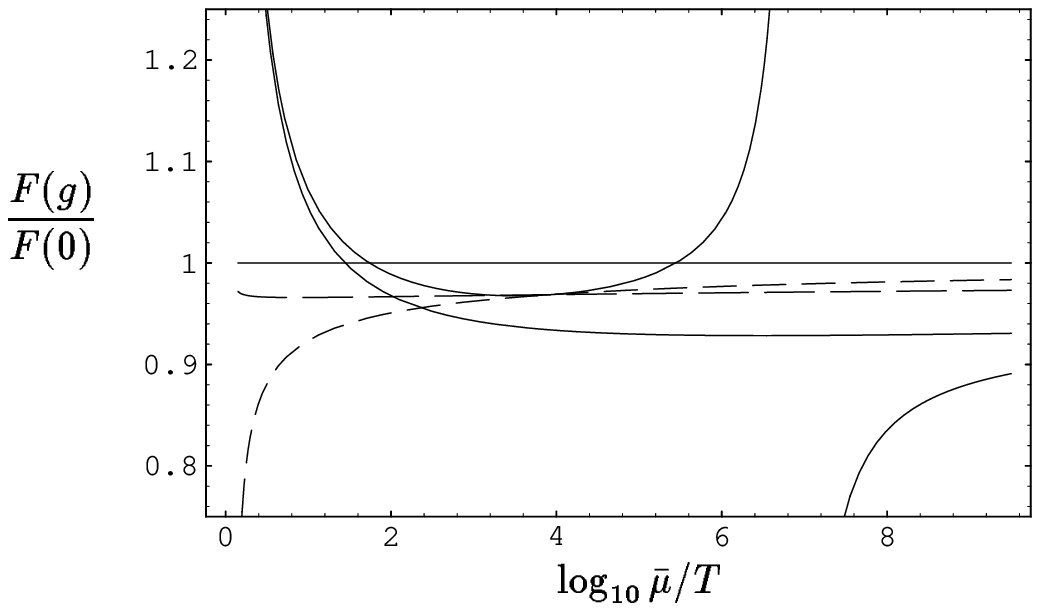,width=5.875in,bbllx=36pt,bblly=482pt,bburx=505pt,bbury=655pt}
\begin{center}
Fig.~4b
\end{center}

\caption{The same as Fig.~1, but for only three fermion flavors and
with $\al(2\pi T)=1/3$.}
\end{figure}

\begin{figure}

\psfig{figure=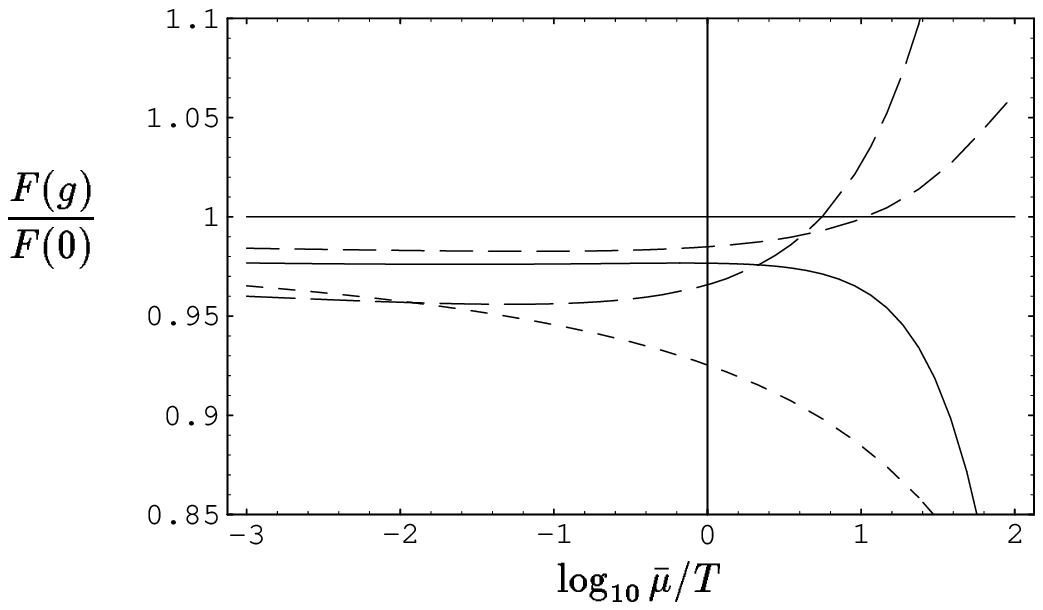,width=5.875in,bbllx=36pt,bblly=482pt,bburx=505pt,bbury=655pt}
\begin{center}
Fig.~5a
\end{center}

\psfig{figure=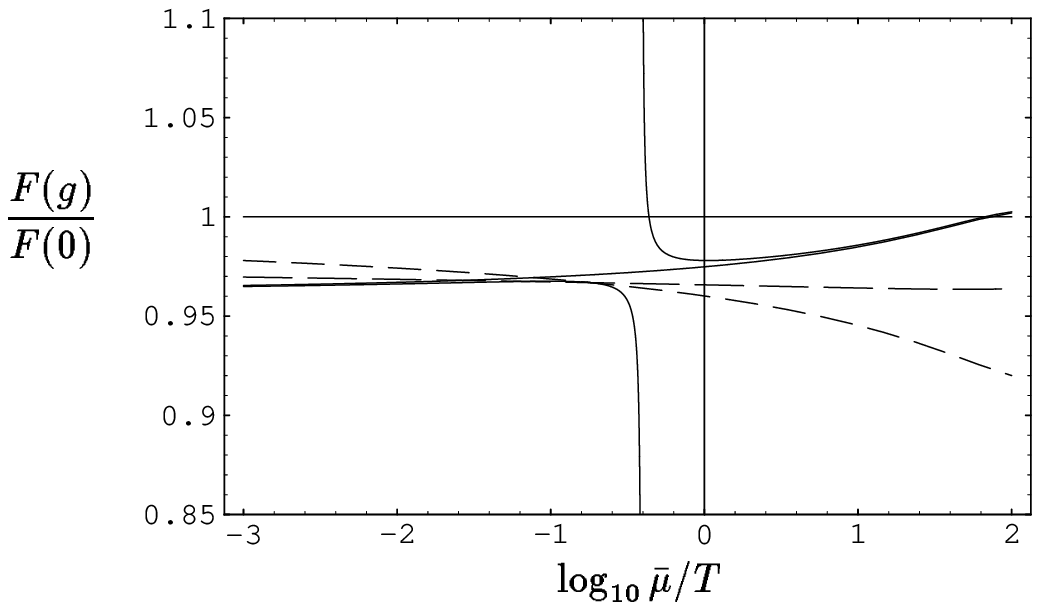,width=5.875in,bbllx=36pt,bblly=482pt,bburx=505pt,bbury=655pt}
\begin{center}
Fig.~5b
\end{center}

\caption{The same as Fig.~1, but in scalar theory with $\al(T)=0.75$.}
\end{figure}

\end{document}